\documentclass[a4paper, oneside, twocolumn, notitlepage, 9pt]{extarticle_ecoc2009}
\usepackage{ecoc2009}

\usepackage{amsmath, amssymb}
\usepackage{graphicx}
\usepackage{subfig}

\usepackage{xspace} 

\usepackage[
	ruled,
	vlined,
	boxed, 
	linesnumbered, 
	commentsnumbered]{algorithm2e}

\usepackage[labelfont=bf, justification=centering]{caption}

\usepackage{enumitem}
\setlist[itemize]{noitemsep, topsep=0pt, itemindent=0pt,
leftmargin=0.7cm}
\setlist[enumerate]{noitemsep, topsep=0pt, itemindent=0pt, 
leftmargin=0.7cm}

\newcommand\xqed[1]{%
\leavevmode\unskip\penalty9999 \hbox{}\nobreak\hfill
\quad\hbox{#1}}
\newcommand\demo{\xqed{$\triangle$}}

\newcommand{\transpose}{\intercal}

\newcommand{\vect}[1]{\ensuremath{\boldsymbol{#1}}}
\newcommand{\mat}[1]{\ensuremath{\mathbf{#1}}}

\begin{document}

\selectlanguage{english}    


\title{Miscorrection-free Decoding of Staircase Codes}

\newcommand{\aff}[1]{\textsuperscript{(#1)}}

\author{
	Christian H\"ager\aff{1,2}
	and
	Henry D.~Pfister\aff{2}
%
}

\maketitle                  

\begin{strip}
  \begin{author_descr}

    \textsuperscript{(1)} Department of Signals and Systems, Chalmers
		University, Sweden, 
    \mailto{christian.haeger@chalmers.se}


    \textsuperscript{(2)} Department of Electrical and Computer
		Engineering, Duke University, US
		


  \end{author_descr}
\end{strip}

\setstretch{1.07}

\begin{strip}
  \begin{ecoc_abstract}
		We propose a novel decoding algorithm for staircase codes which
		reduces the effect of undetected component code miscorrections.
		The algorithm significantly improves performance, while
		retaining a low-complexity implementation suitable for high-speed
		optical transport networks. 
		
  \end{ecoc_abstract}
\end{strip}


\section{Introduction}

Hard-decision forward error correction (HD-FEC) can offer dramatically
reduced complexity compared to soft-decision FEC, at the price of some
performance loss. HD-FEC is used, for example, in regional/metro
optical transport networks (OTNs)\cite{Justesen2010} and has also been
considered for other cost-sensitive applications such as optical data
center interconnects\cite{Yu2017}. Our focus is on staircase
codes\cite{Smith2012a}, which provide excellent performance and have
received considerable attention in the literature. 




Similar to classical product codes, staircase codes are built from
short component codes and decoded by iteratively applying
bounded-distance decoding (BDD) to the component codes. For the
purpose of this paper, BDD of a $t$-error-correcting component code
can be seen as a black box that operates as follows.  Let $\vect{r} =
\vect{c} + \vect{e}$, where $\vect{c}, \vect{e} \in \{0,1\}^n$ denote
a component codeword and random error vector, respectively, and $n$ is
the code length. BDD yields the correct codeword $\vect{c}$ if
$d_\text{H}(\vect{r}, \vect{c}) = w_\text{H}(\vect{e}) \leq t$, where
$d_\text{H}$ and $w_\text{H}$ denote Hamming distance and weight,
respectively. On the other hand, if $w_\text{H}(\vect{e}) > t$, the
decoding either fails or there exists another codeword $\vect{c}'$
such that $d_\text{H}(\vect{r},\vect{c}') \leq t$. In the latter case,
BDD is technically successful but the decoded codeword $\vect{c}'$ is
not the correct one. Such \emph{miscorrections} are highly undesirable
because they introduce additional errors into the iterative decoding
process and significantly degrade performance.

In this paper, we propose a novel iterative HD decoding algorithm for
staircase codes which can detect and avoid most miscorrections. The
algorithm provides significant post-FEC bit error rate improvements,
in particular when $t$ is small (which is typically the case in
practice). As an example, for $t=2$, the algorithm can improve
performance by roughly $0.4\,$dB and reduce the error floor by over an
order of magnitude, up to the point where the iterative decoding
process is virtually miscorrection-free. Error floor improvements are
particularly important for applications with stringent reliability
constraints such as OTNs. 


\vspace{-0.07cm}

\section{Staircase codes and iterative decoding}


Let $\mathcal{C}$ be a binary linear component code with length $n$
and dimension $k$. Assuming that $n$ is even, a staircase code with
rate $R = 2 k/n - 1$ based on $\mathcal{C}$ is defined as the set of
all matrix sequences $\mat{B}_k \in \{0,1\}^{a \times a}$, $k =
0,1,2,\dots$, such that the rows in $[\mat{B}^\transpose_{k-1},
\mat{B}_k]$ for all $k \geq 1$ form valid codewords of $\mathcal{C}$,
where $a = n/2$ is the block size and $\mat{B}_0$ is the all-zero
matrix. 

We use extended primitive Bose--Chaudhuri--Hocquenghem (BCH) codes as
component codes, i.e., a BCH code with an additional parity bit formed
by adding (modulo 2) all $2^{\nu}-1$ coded bits of the BCH code, where
$\nu$ is the Galois field extension degree. The overall extended code
then has length $n = 2^{\nu}$ and guaranteed dimension $k = 2^\nu-\nu
t-1$.  The extra parity bit increases the guaranteed minimum distance
to $d_\text{min} = 2t+2$. 



\begin{figure}[t]
	\begin{center}
		\includegraphics[width=7.33cm]{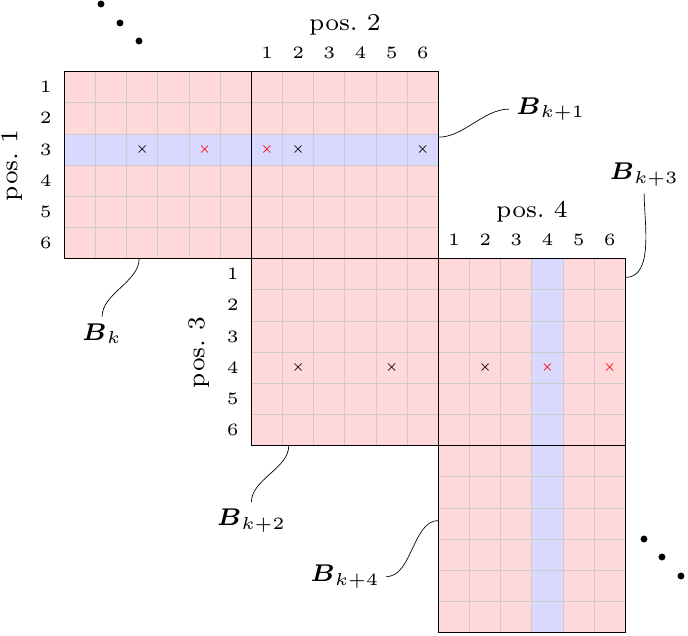}
	\end{center}
	\caption{Staircase decoding window of size $W=5$}
	\label{fig:staircase_array}
\end{figure}

The conventional decoding procedure for staircase codes uses a sliding
window comprising $W$ received blocks $\mat{B}_k, \mat{B}_{k+1},
\dots, \mat{B}_{k+W-1}$. This is illustrated in
Fig.~\ref{fig:staircase_array} for $W = 5$ and $a = 6$.  It is
convenient to identify each component code in the window by a tuple
$(i,j)$, where $i \in \{1, 2, \dots, W-1\}$ indicates the position
relative to the current decoding window and $j \in \{1, 2, \dots, a\}$
enumerates all codes at a particular position. As an example, the
component codes $(1,3)$ and $(4,4)$ are highlighted in blue in
Fig.~\ref{fig:staircase_array}. Pseudocode for the conventional
decoding procedure is given in Algorithm 1 below. Essentially, all
component codes are decoded $\ell$ times, after which the decoding
window shifts to the next position. Note that after the window shifts,
the same component code is identified by a different position index. 


\setlength{\textfloatsep}{10pt}
\begin{algorithm}[b]
	\small
	\DontPrintSemicolon
	\SetKw{ShortFor}{for}
	\SetKw{Break}{break}
	\SetKw{MyWhile}{while}
	\SetKw{MyIf}{if}
	\SetKw{MySet}{set}
	\SetKw{MyElse}{else}
	\SetKw{MyCompute}{compute}
	\SetKw{KwEach}{each}
	\SetKw{KwAnd}{and}

	$k \leftarrow 0$\;
	\While{true}{
		\For{$l = 1, 2, \dots, \ell$}{
			\For{$i = W, W-1, \dots, 1$}{
				\For{$j = 1, 2, \dots, a$}{
					apply BDD to component code $(i,j)$
				}
			}
		}
		output decision for $\mat{B}_k$ and shift window\;
		$k \leftarrow k + 1$\;
	}
	\caption{ {\small Window decoding of staircase codes} }
\end{algorithm}



\section{Performance analysis}

Analyzing the post-FEC bit error rate of staircase codes under the
conventional decoding procedure is challenging. A major simplification
is obtained by assuming that no miscorrections occur in the BDD of the
component codes. In this case, it is possible to rigorously
characterize the asymptotic performance as $a \to \infty$ using a
technique called density evolution\cite{Haeger2017tit}. Moreover, the
error floor can be estimated by enumerating stopping sets, also known
as stall patterns\cite{Smith2012a}.  However, if miscorrections are
taken into account, both the asymptotic and error floor predictions
are nonrigorous and become inaccurate. 

\setlength{\textfloatsep}{20pt}

\begin{figure}[t]
	\begin{center}
		\includegraphics[width=7.4cm]{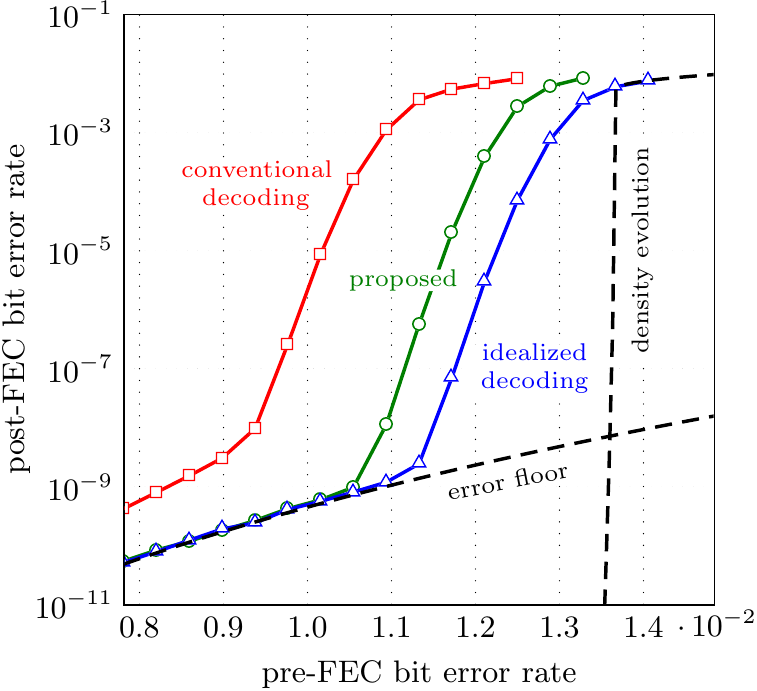}
	\end{center}
	\caption{Results for component codes with $n=256$ and $t=2$}
	\label{fig:scc_results}
\end{figure}

\emph{Example 1:} Let $\nu=8$ and $t=2$, which gives a staircase code
with $a = 128$ and $R = 0.867$. For window decoding parameters $W = 8$
and $\ell = 7$, the density evolution and error floor predictions are
shown in Fig.~\ref{fig:scc_results} by the dashed lines. The analysis
can be verified by performing idealized decoding, where miscorrections
are prevented during BDD. The results are shown by the blue line
(triangles) in Fig.~\ref{fig:scc_results} and accurately match the
theoretical predictions. However, the actual performance with true BDD
deviates from the idealized decoding, as shown by the red line
(squares). \demo

The performance degradation with respect to idealized decoding becomes
less severe for larger values of $t$. Unfortunately, small values of
$t$ are commonly used in practice because BDD can be implemented very
efficiently in this case.  We note at this point that there exist
several works that attempt to quantify the performance loss due to
miscorrections. In terms of error floor, the work in \cite{Smith2012a}
introduces a heuristic parameter, whose value unfortunately has to be
estimated from simulative data. In terms of asymptotic performance,
the authors are aware of two works\cite{Jian2015,Truhachev2016}, both
of which do not directly apply to staircase codes, but to a related
code ensemble. 


\section{Proposed algorithm}


The main idea in order to improve performance is to systematically
exploit the fact that miscorrections lead to inconsistencies, in the
sense that two component codes that protect the same bit may disagree
on its value. In the following, we show how these inconsistencies can
be used to (a) reliably prevent miscorrections and (b) identify
miscorrected codewords in order to revert their decoding decisions. 

Our algorithm relies on so-called anchor codewords, which have
presumably been decoded without miscorrections. Roughly speaking, we
want to make sure that bit flips do not lead to inconsistencies with
anchor codewords. Consequently, decoding decisions from codewords that
are in conflict with anchors are not applied. However, a small number
of anchor codewords may be miscorrected and we allow for the decoding
decisions of anchors to be reverted if too many other codewords are in
conflict with a particular anchor. 

In order to make this more precise, we regard the BDD of a component
code $(i,j)$ as a two-step process. In the first step, the decoding is
performed and the outcome is either a set of error locations
$\mathcal{E}_{i,j} \subset \{1, 2, \dots, n\}$, where
$|\mathcal{E}_{i,j}| \leq t$, or a decoding failure. In the second
step, error-correction is performed by flipping the bits corresponding
to the error locations. Initially, we only perform the decoding step
for all component codes, i.e., all component codes in the decoding
window are decoded without applying any bit flips. We then iterate
$\ell$ times over the component codes in the same fashion as in
Algorithm 1, but replacing line 6 with the following four steps:  

\begin{enumerate}
	\item If no decoding failure occurred for the component codeword
		$(i,j)$, we proceed to step 2, otherwise, we skip to the next
		codeword.

	\item For each $e \in \mathcal{E}_{i,j}$, check if $e$ corresponds
		to an anchor codeword. If so, let $C$ be the number of other
		conflicts that this anchor is involved in. If $C < T$, where $T$
		is a fixed threshold, the codeword $(i,j)$ is frozen and we skip
		to the next codeword. Frozen codewords are always skipped (in the
		loop of Algorithm 1) for the rest of the decoding unless any of
		their bits change. If $C \geq T$, the anchor is marked for
		backtracking. 

	\item Error-correction for codeword $(i,j)$ is applied, i.e., the
		bits at all error locations in $\mathcal{E}_{i,j}$ are flipped. We
		also apply the decoding step again for codewords that had their
		syndrome changed due to a bit flip. Finally, the codeword $(i,j)$
		becomes an anchor. 
		
	\item Lastly, previously applied bit flips are reversed for all
		anchor codewords that were marked for backtracking during step 2.
		These codewords are no longer anchors and all frozen codewords that
		were in conflict with these codewords are unfrozen. 
		
\end{enumerate}

Note that steps 3 and 4 are not reached for codeword $(i,j)$ if the
corresponding bit flips of that codeword are inconsistent with
any anchor for which $C < T$ holds. 

The following two examples illustrate the above steps for $t=2$ and
$T=1$ with the help of Fig.~\ref{fig:staircase_array}.

\emph{Example 2:} Assume that we are at $(i,j) = (3,4)$, corresponding
to a component code with three attached errors shown by the black
crosses. The codeword is miscorrected with $\mathcal{E}_{3,4} = \{10,
12\}$ shown by the red crosses. Assuming that the codeword $(4,4)$ is
an anchor without any other conflicts, the codeword $(3,4)$ is frozen
during step 2 and no bit flips are applied. \demo

\emph{Example 3:} Let the codeword $(1,3)$ in
Fig.~\ref{fig:staircase_array} be a miscorrected anchor without
conflicts and error locations $\mathcal{E}_{1,3} = \{5, 7\}$. Assume
that we are at $(i,j) = (2,1)$. The codeword $(2,1)$ has one attached
error, thus $\mathcal{E}_{2,1} = \{3\}$. During step 2, the codeword
$(2,1)$ is frozen and we skip to codeword $(2,2)$ with
$\mathcal{E}_{2,2} = \{3, 10\}$. The bit flip at $e = 3$ is
inconsistent with the anchor $(1,3)$, but, since this anchor is
already in conflict with $(2,1)$ (and, hence, $C = T = 1$), the anchor
is marked for backtracking.  The bits according to $\mathcal{E}_{2,2}$
are then flipped in step 3 and the anchor $(1,3)$ is backtracked in
step 4. \demo

The previous example shows how a miscorrected anchor is
backtracked. Since we do not know if an anchor is
miscorrected or not, it is also possible that we mistakenly backtrack
``good'' anchors. Fortunately, this is unlikely to happen for long
component codes because the additional errors due to miscorrections
are approximately randomly distributed within the
codeword. This implies that errors of two (or more)
miscorrected codewords rarely overlap. 



For the algorithm to work well, a sufficiently large fraction of
codewords at each position should be ``good'' anchors.  However, when
the decoding window shifts and a new block is added, no anchors exist
at the last position $W-1$. We found that it is therefore beneficial
to artificially restrict the error-correcting capability of these
component codes in order to avoid anchoring too many miscorrected
codewords. For example, for $t=2$, all component codes at position
$W-1$ are treated as single-error-correcting.  This restriction
reduces the probability of miscorrecting a component code by roughly a
factor of $n$, which is significant for long component
codes\cite{Justesen2011}. Note that due to the window decoding, we are
merely gradually increasing the error-correction capability: once the
decoding window shifts, the component codes shift as well and they are
then decoded with their full error-correcting capability. 

We remark that essentially the same algorithm can also be applied to
product codes and other related code constructions, e.g.,
half-product or braided codes.


\section{Decoding complexity}

In terms of decoding complexity, one of the main advantages of
iterative HD decoding of staircase codes compared to message-passing
decoding of LDPC codes is the significantly reduced decoder data flow
requirement\cite{Smith2012a}. While a thorough complexity analysis for
the proposed algorithm is beyond the scope of this paper, we note that
the algorithm can operate entirely in the syndrome domain, thereby
leveraging the syndrome compression effect that is described
in\cite{Smith2012a}. However, additional storage is needed compared to
the conventional decoding to keep track of the error locations of
anchor codewords (in case they are backtracked) and to store the
conflicts between codewords. 


\section{Results and Discussion}


We consider the same parameters as in Example 1, i.e., $\nu=8$, $t=2$,
$W=8$, and $\ell= 7$. The conflict threshold is set to $T=1$ and we apply
the error-correction capability restriction for component codes at
position $W-1$ as described above.
Simulation results for the proposed algorithm are shown in
Fig.~\ref{fig:scc_results} by the green line (circles). It can be seen
that the performance is significantly improved compared to the
conventional decoding.
In particular in terms of the error floor, the performance is
virtually identical to the idealized decoding where miscorrections are
prevented.  Overall, the improvements translate into an additional
coding gain of around $0.4\,$dB at a post-FEC bit error rate of
$10^{-9}$ over the conventional decoding.

Note that the staircase code parameters were chosen such that the
error floor is high enough to be within the reach of software
simulations. For OTNs, post-FEC bit error rates below $10^{-15}$ are
typically required. In this case, other code parameters should be used
or one may apply post-processing techniques to reduce the error floor
below the application requirements\cite{Holzbaur2017}. 


\section{Conclusion}

We have shown that the post-FEC performance of staircase codes can be
significantly improved by adopting a modified iterative HD decoding
algorithm that reduces the effect of miscorrections. For component
codes with error-correcting capability $t = 2$, an additional coding
gain of around $0.4\,$dB can be achieved. Moreover, the error floor
can be reduced by over an order of magnitude, giving virtually
miscorrection-free performance. 

\vspace{-0.1cm}

\section{Acknowledgements}

{\footnotesize 

This work is part of a project that has received
funding from the European Union's Horizon 2020 research and innovation
programme under the Marie Sk\l{}odowska-Curie grant agreement
No.~749798. The work was also supported in part by the National
Science Foundation (NSF) under Grant No.~1609327. Any opinions,
findings, recommendations, and conclusions expressed in this material
are those of the authors and do not necessarily reflect the views of
these sponsors.

}


\vspace{-0.1cm}

\newif\iffullbib
\fullbibfalse


\newlength{\bibspace}
\setlength\bibspace{-2.5mm}

\newcommand{\jlt}{J.~Lightw.~Technol.}
\newcommand{\ope}{Opt.~Exp.}
\newcommand{\tit}{IEEE Trans.~Inf.~Theory}
\newcommand{\tc}{IEEE Trans.~Comm.}
\newcommand{\ofc}{Proc.~OFC}
\newcommand{\ecoc}{Proc.~ECOC}
\newcommand{\ita}{Proc.~ITA}
\newcommand{\scc}{Proc.~SCC}

\iffullbib

{\scriptsize
\setlength{\bibsep}{0.3ex plus 0.3ex}
\bibliographystyle{IEEEtran}%
\bibliography{$HOME/lib/bibtex/library_mendeley}%
}%

\else

{\scriptsize
\setlength{\bibsep}{0.3ex plus 0.3ex}

}

\fi

\end{document}